\documentclass[journal,comsoc]{IEEEtran}
\usepackage[ruled,vlined,linesnumbered]{algorithm2e}
\usepackage{algorithmic}
\newcommand{\note}[1]{}

\usepackage{amsthm}

\usepackage{xcolor, soul}
\sethlcolor{yellow}
\usepackage{xcolor}
\usepackage{balance}

\usepackage{mathrsfs}  
\usepackage{graphicx}
\usepackage{xcolor}
\usepackage{graphicx,epsf,epic,eepic,psfig,latexcad}
\renewcommand{\baselinestretch}{1}

\usepackage{subcaption}
\usepackage{amssymb,amsmath,caption}
\usepackage{cite}
\usepackage{amsmath, amssymb, amsthm}
\usepackage{graphicx}
\renewcommand{\baselinestretch}{1}
\usepackage{epstopdf}
\makeatletter
\usepackage{textcomp}
\makeatother
\usepackage{xcolor}
\begin{document}
	
	\title{MAP-Based Task-Oriented Precoding for Multiuser Communication}
\author{Mohammad Javad Ahmadi, Zhentian Zhang~\IEEEmembership{Member,~IEEE}, Rafael F. Schaefer~\IEEEmembership{Senior Member,~IEEE},\\
    and H. Vincent Poor,~\IEEEmembership{Life Fellow,~IEEE}
    \thanks{This work of M. J. Ahmadi and R. F. Schaefer have been supported in part by the German Federal Ministry of Research, Technology and Space (BMFTR) through the transfer hub \emph{6G-life} under Grant 16KIS2413K and in part by the German Research Foundation (DFG) as part of Germany's Excellence Strategy EXC 2050/2 -- Project ID 390696704 -- Cluster of Excellence \emph{``Centre for Tactile Internet with Human-in-the-Loop'' (CeTI)} and within the Priority Programme SPP 2378 ``Resilient Worlds'' -- Project ID 503657103. The work of H. V. Poor was supported in part by an Innovation Grant from Princeton NextG.}
\thanks{M. J. Ahmadi and R. F. Schaefer are with the Chair of Information Theory and Machine Learning, Technische Universität Dresden, 01062 Dresden, Germany, and also with the Cluster of Excellence ``CeTI'', Technische Universität Dresden, 01062 Dresden, Germany (e-mail: \{mohammad\_javad.ahmadi, rafael.schaefer\}@tu-dresden.de).}

\thanks{Z. Zhang is with The Hong Kong Polytechnic University, Hong Kong SAR, China (e-mail: zhentianzhangzzt@gmail.com).}

\thanks{H. V. Poor is with the Department of Electrical and Computer Engineering, Princeton University (e-mail: poor@princeton.edu).}
}
	    \vspace{-33mm}
	\maketitle
	\begin{abstract}
We propose a task-oriented multiuser wireless communication framework for distributed classification based on a MAP-driven system design under wireless channel impairments. \textcolor{black}{By deriving tractable approximations of the MAP error bound}, the proposed approach enables the design of learning-based feature extraction and precoding strategies. Unlike existing approaches that optimize intermediate reconstruction, information-theoretic, or feature-separability objectives, the proposed formulation directly optimizes objectives derived from the MAP decision error, thereby providing a more direct connection to the final classification performance. Simulation results demonstrate that the proposed schemes achieve higher classification accuracy than their counterparts, with lower or comparable computational complexity.
\end{abstract}
\begin{IEEEkeywords}
Task-oriented communication, MAP detection, feature learning, wireless precoding, ModelNet10.
\end{IEEEkeywords}

\vspace{-2mm}
\section{Introduction}
Task-oriented and semantic communication have recently emerged as a promising paradigm for next-generation wireless systems, where the goal is to reliably convey task-relevant semantic information rather than reconstructing raw transmitted signals~\cite{Mostaani2022Task}. In contrast to conventional communication systems that primarily focus on minimizing symbol or signal distortion, semantic and task-oriented frameworks aim to optimize end-task performance such as classification or inference accuracy at the receiver. This shift has motivated the integration of machine learning techniques with wireless communication design, enabling joint optimization of feature representation, transmission strategies, and decision-making under channel impairments. As a result, new system designs are required that account for both learning-based feature extraction and channel-aware signal processing in a unified framework~\cite{bourtsoulatze2019deepjscc,jankowski2021retrieval,shao2022ib,xie2021semantic,LMMSE,Cai2024MCR2,Cai2025MIMO}.

Earlier works in task-oriented communication focus on learning-based feature transmission, where neural networks extract compact representations and transmit them over wireless channels for downstream tasks such as classification or retrieval~\cite{bourtsoulatze2019deepjscc,jankowski2021retrieval,shao2022ib}. These methods mainly optimize task performance indirectly through representation learning frameworks such as deep joint source–channel coding (JSCC) or information bottleneck models. More recent approaches further shape the learned feature space to enhance post-channel inference performance, rather than solely compressing representations for transmission~\cite{Cai2025MIMO,Cai2024MCR2,LMMSE,xie2021semantic,Wen2024TSCCI}. \textcolor{black}{Although achieving promising performance, these methods employ different task-oriented objectives that are related to, but not directly derived from, the final inference error. For instance,~\cite{LMMSE} targets accurate signal reconstruction via the linear minimum mean-squared error (LMMSE) criterion, while~\cite{xie2021semantic,Cai2025MIMO} employ mutual information as the optimization objective. The method in~\cite{Cai2024MCR2} employs maximal coding rate reduction (MCR$^2$) to enhance class separability in the learned feature space, while~\cite{Wen2024TSCCI} employs discriminant gain to characterize class separability. Although these objectives are related to the ultimate inference task, they characterize task performance through intermediate reconstruction, information-theoretic, or distributional separability metrics.}

\textcolor{black}{To directly characterize the final inference performance, this work develops a MAP-based formulation and derives tractable surrogate objectives for feature extraction and precoding based on the MAP decision error probability.} By approximating the MAP error, we obtain tractable objectives that reduce the computational complexity by eliminating the need for covariance matrix inversions and eigenvalue decompositions. Simulation results demonstrate improved classification accuracy and reduced computational time compared to the state of the art.

The rest of this letter is organized as follows. Section~\ref{Secion_SysModel} introduces the system model. Section~\ref{SectionMain} first presents the MAP detection rule and derives the corresponding error probability, and then develops the proposed feature extraction and precoding strategies \textcolor{black}{using tractable surrogate objectives derived from the MAP error formulation.} Section~\ref{Sec_numResults} provides numerical results to validate the proposed method. Finally, Section~\ref{SecConclusion} concludes the letter.

\section{System Model}
\label{Secion_SysModel}
We consider a distributed wireless classification system consisting of $K$ edge devices (workers) and a central server (fusion center). Each worker $k \in \{1,\dots,K\}$ is equipped with $N_k$ transmit antennas, while the server is equipped with $M$ receive antennas. The observed data at each worker belongs to one of $C$ classes $\{\mathcal{C}_1,\dots,\mathcal{C}_C\}$. The $k$th worker processes its high-dimensional observation $\mathbf{s}_k$ using a feature extraction function $\mathbf{x}_k = f_{\theta_k}(\mathbf{s}_k)\in \mathbb{C}^{D_k \times 1}$, where $f_{\theta_k}(\cdot)$ denotes a parameterized nonlinear mapping (e.g., a neural network with parameters $\theta$) that transforms the input sample into a lower-dimensional task-relevant feature vector. The resulting feature vector $\mathbf{x}_k$ is transmitted to the server through a wireless channel. The ultimate goal of the system is to correctly infer the class label $\mathcal{C}_j$ from the received signal at the fusion center. The received signal at the edge server corresponding to the $t$th channel use is given by
\begin{align}
\mathbf{y}_t &= \sum_{k=1}^K \mathbf{H}_k \mathbf{V}_{k,t} \mathbf{x}_k + \mathbf{z}_t,\label{ReceivedSignal1}
\end{align}
where $\mathbf{H}_k \in \mathbb{C}^{M \times N_k}$ is the wireless channel between worker $k$ and the server, $\mathbf{V}_{k,t} \in \mathbb{C}^{N_k \times D_k}$ is the precoding matrix at the $t$th channel-use of the $k$th worker, and $\mathbf{z}_t\sim \mathcal{CN}(\mathbf{0},\sigma^2\mathbf{I}_{M})$ is the circularly symmetric complex Gaussian noise. Stacking the received signals over \(T\) channel uses, we can write
\begin{align}
    \mathbf{y}&=\sum_{k=1}^K\tilde{\mathbf{H}}_k\mathbf{V}_k\mathbf{x}_k+\mathbf{z},\label{Eq_receivedSignal23}
\end{align}
where $\tilde{\mathbf{H}}_k=\mathbf{I}_T\otimes \mathbf{H}_k\in \mathbb{C}^{TM\times TN_k}$, \ $\mathbf{y}=[\mathbf{y}_1^T,\mathbf{y}_2^T,....,\mathbf{y}_T^T]^T\in \mathbb{C}^{TM\times 1}$, $\mathbf{z}=[\mathbf{z}_1^T,\mathbf{z}_2^T,....,\mathbf{z}_T^T]^T\in \mathbb{C}^{TM\times 1}$, $\mathbf{V}_k=\left[\mathbf{V}_{k,1}^T, \mathbf{V}_{k,2}^T, \dots, \mathbf{V}_{k,T}^T\right]^T\in \mathbb{C}^{TN_k\times D_k}$.

The received signal in \eqref{Eq_receivedSignal23} can equivalently be written as
\begin{align}
    \mathbf{y}&=\tilde{\mathbf{H}}\mathbf{V}\mathbf{x}+\mathbf{z},\label{Eq_receivedSignal24}
\end{align}
where $\tilde{\mathbf{H}}=[\tilde{\mathbf{H}}_1,\tilde{\mathbf{H}}_2,...,\tilde{\mathbf{H}}_K]\in \mathbb{C}^{MT\times TN}$, $\mathbf{x}=[\mathbf{x}_1^T,\mathbf{x}_2^T,\ldots,\mathbf{x}_K^T]^T
\in\mathbb{C}^{D\times1}$, and $\mathbf{V}=\mathrm{blkdiag}(\mathbf{V}_{1}, \mathbf{V}_{2}, \dots, \mathbf{V}_{K})\in\mathbb{C}^{TN\times D}$ with $N=\sum_{k=1}^KN_k$ and $D=\sum_{k=1}^KD_k$. In \eqref{Eq_receivedSignal24}, $\tilde{\mathbf{H}}$ denotes the effective channel matrix and $\mathbf{V}\mathbf{x}$ is the transmitted signal. To satisfy the transmit power constraint, we impose
\begin{align}
    \mathbb{E}\{\|\mathbf{V}\mathbf{x}\|^2\} \leq PT, \label{eq_powerConstraint}
\end{align}
where $P$ denotes the total transmit power budget across all users.

\section{MAP-Based Task-Oriented System Design}
\label{SectionMain}
In this section, we formulate a MAP-based task-oriented design framework. First, in Section~\ref{MAPError}, we present the MAP detection rule at the receiver side and derive the corresponding error probability. Then, in Sections~\ref{SectionNN} and \ref{Section_prcoder}, we develop feature-extraction and precoding methods based on \textcolor{black}{tractable approximations of the MAP error}.

\subsection{MAP Detection and Error Probability Analysis}
\label{MAPError}

Assuming class-conditional Gaussian feature distributions, i.e.,
$\mathbf{x}|\mathcal{C}_j\sim \mathcal{CN}(\mu_j,\Sigma_j)$, and according to the received signal model in \eqref{Eq_receivedSignal24}, the received signal conditioned on class $\mathcal{C}_j$ follows a complex Gaussian distribution given by $\mathbf{y}|\mathcal{C}_j
\sim
\mathcal{CN}(\boldsymbol{\mu}_j,\mathbf{K}_j)$, where $\boldsymbol{\mu}_j=\tilde{\mathbf{H}}\mathbf{V}\mu_j$ and $\mathbf{K}_j=\tilde{\mathbf{H}}\mathbf{V}\Sigma_j\mathbf{V}^{H}\tilde{\mathbf{H}}^{H}
+\sigma^2\mathbf{I}_{MT}$.

The MAP detector at the receiver side can be expressed as 
$\hat{c}
=
\arg\max_j
\mathbb{P}(\mathbf{y}|\mathcal{C}_j)p_j$,
where $p_j=\mathbb{P}(\mathcal{C}_j)$ denotes the class prior probability. Due to the Gaussian distributions of $\mathbf{y}$, the MAP detector can be written as
\begin{align}
\hat{c}
=
\arg\min_j
\Big\{
\|\mathbf{y}&-\boldsymbol{\mu}_j\|^2_{
\mathbf{K}_j^{-1}}+
\log(|{\mathbf{K}_j}|/{p_j})
\Big\},
\label{MAP_general_form}
\end{align}
where $\|\mathbf{a}\|_{\mathbf{A}}^2\triangleq\mathbf{a}^H\mathbf{A}\mathbf{a}$.

The MAP detection error probability can be upper-bounded using a pairwise union bound as
\begin{align}
\textcolor{black}{P_e \leq \sum_{j=1}^C p_j\sum_{k\neq j}  P_{jk}},\label{errorTermNew}
\end{align}
where $P_{jk}$ denotes the pairwise error probability of deciding class $\mathcal{C}_k$ when the true class is $\mathcal{C}_j$, which is given by
\begin{align}
P_{jk}
=
\mathbb{P}\Bigg(
\|\mathbf y-\boldsymbol{\mu}_{k}\|^2_{\mathbf{K}_k^{-1}}+a_{jk} <
\|\mathbf y-\boldsymbol{\mu}_j\|^2_{\mathbf{K}_j^{-1}}
\Bigg),
\end{align}
where $a_{jk}
=
\log
\left(
\frac{
|\mathbf{K}_k|p_j
}{
|\mathbf{K}_j|p_k
}
\right)$.
Letting $\mathbf{y}|\mathcal{C}_j=\boldsymbol{\mu}_j+\mathbf{k}_j$ with $\mathbf{k}_j\sim\mathcal{CN}(\mathbf{0},\mathbf{K}_j)$, the pairwise error probability can be expressed as
\begin{align}
P_{jk}
&=
\mathbb{P}\Bigg(n_{jk}>
\|\boldsymbol{\mu}_{jk}\|_{\mathbf{K}_k^{-1}}^2
+
b_{jk}
+
a_{jk}
\Bigg)
\nonumber\\
&=
\mathbb{E}_{b_{jk}}
\left\{
\textcolor{black}{Q\left(
\frac{
\|\boldsymbol{\mu}_{jk}\|_{\mathbf{K}_k^{-1}}^2
+
b_{jk}
+
a_{jk}
}
{
\sqrt{
2\,
\|\boldsymbol{\mu}_{jk}\|_
{\mathbf{K}_k^{-1}
\mathbf{K}_j
\mathbf{K}_k^{-1}}^2
}
}
\right)}
\right\},
\label{eq_pairwiseExpanded}
\end{align}
where $\boldsymbol{\mu}_{jk}=\boldsymbol{\mu}_j-\boldsymbol{\mu}_{k}$, $n_{jk}=-2\Re\!\left\{
\mathbf{k}_j^{H}
\mathbf{K}_k^{-1}
\boldsymbol{\mu}_{jk}
\right\}$ and $b_{jk}
=
\mathbf{k}_j^{H}
\left(
\mathbf{K}_k^{-1}
-
\mathbf{K}_j^{-1}
\right)
\mathbf{k}_j$.

\subsection{Neural Network Training and Feature Learning}
\label{SectionNN}
In this part, we describe the feature learning module used to generate class-discriminative representations at each user. Specifically, the $k$th worker employs a neural network parameterized by $\theta_k$, denoted by $f_{\theta_k}(\cdot)$, to map its high-dimensional observation into a real-valued latent representation $\tilde{\mathbf{x}}_k \in \mathbb{R}^{2D_k\times 1}$. The resulting vector is converted into a complex representation $\mathbf{x}_k \in \mathbb{C}^{D_k \times 1}$ as $\mathbf{x}_k = \tilde{\mathbf{x}}_{k,1:D_k} + j \tilde{\mathbf{x}}_{k,D_k+1:2D_k}$, followed by normalization such that $\|\mathbf{x}_k\|^2 = 1$. The feature vectors from $K$ workers are then aggregated as $\mathbf{x}=[\mathbf{x}_1^T,\ldots,\mathbf{x}_K^T]^T\in \mathbb{C}^{D\times 1}$. The feature vectors from the $K$ workers are then aggregated as
$\mathbf{x}=[\mathbf{x}_1^T,\ldots,\mathbf{x}_K^T]^T\in\mathbb{C}^{D\times1}$.
During training, the class means $\{\boldsymbol{\mu}_j\}_{j=1}^C$ of the aggregated feature representation are computed from each mini-batch and incorporated into the objective function. The neural network parameters $\{\theta_k\}_{k=1}^K$ are jointly optimized using standard backpropagation. After training, the class means and covariance matrices are estimated from the learned feature representations and subsequently used for precoder design. The training procedure is summarized in Algorithm~\ref{alg_feature}, where $N_s$ denotes the number of training samples. The proposed loss function is described below.

\begin{algorithm}
\caption{Training Feature Extractor}
\label{alg_feature}
\textcolor{black}{
{\small
\textbf{Input:} Training dataset, batch size $B$, initialized network parameters $\{\theta_k\}_{k=1}^{K}$.\\
\textbf{Output:} Trained feature extractors $\{f_{\theta_k}(\cdot)\}_{k=1}^{K}$, class means $\{\mu_j\}_{j=1}^{C}$, and covariance matrices $\{\Sigma_j\}_{j=1}^{C}$.\\
\Repeat{Convergence of $\{\theta_k\}_{k=1}^{K}$}{
\For{$k=1,\ldots,K$}{
 Randomly select $B$ samples for the $k$th worker.\\
Compute $\tilde{\mathbf{x}}_k=f_{\theta_k}(\mathbf{s}_k)\in \mathbb{R}^{2D_k \times 1}$.\\
Complexification: $\mathbf{x}_k = \tilde{\mathbf{x}}_{k,1:D_k} + j\tilde{\mathbf{x}}_{k,D_k+1:2D_k}$.\\
Normalization: $\|\mathbf{x}_k\|^2 = 1$.
}
Aggregation: $\mathbf{x}=[\mathbf{x}_1^T,\ldots,\mathbf{x}_K^T]^T\in \mathbb{C}^{D\times 1}$.\\
Compute mini-batch class means $\{\mu_j\}_{j=1}^{C}$.\\
Compute the objective function in \eqref{eq_channel_independent_Q}.\\
Update $\{\theta_k\}_{k=1}^{K}$ using backpropagation.\\
}
Compute the final class statistics $\{\mu_j,\Sigma_j\}_{j=1}^{C}$ using the trained feature extractors over the entire training set.
}
}
\end{algorithm}

The pairwise error in \eqref{eq_pairwiseExpanded}, and consequently the union bound in \eqref{errorTermNew}, depends on the effective channel-precoder transformation $\mathbf{G}=\tilde{\mathbf{H}}\mathbf{V}$. Hence, the error formulation in \eqref{errorTermNew} cannot be directly employed as the feature-extraction objective, since $\mathbf{G}$ is generally unavailable during feature-extractor training. To obtain a tractable training objective without requiring instantaneous channel state information or the precoder, we statistically characterize $\mathbf{G}$. Specifically, we assume that $\mathbf G$ is zero mean, with uncorrelated elements each having a common second-order moment $\gamma_G$, where $\gamma_G$ characterizes the average effective channel-precoder gain. We further assume that the traces of the class-dependent feature covariance matrices are approximately equal and approximate $\mathrm{tr}(\Sigma_j)\approx\mathrm{tr}(\Sigma)$ for all $j$, where $\mathrm{tr}(\Sigma) := \sum_{l=1}^Cp_j\mathrm{tr}(\Sigma_l)$. Thus, we approximate
 $\mathbf{K}_j\approx \mathbb{E}_{\mathbf{G}}
\left\{
\mathbf{G}\Sigma_j\mathbf{G}^{H}
+\sigma^2\mathbf{I}_{MT}\right\}
=
\delta_H
\mathbf{I}_{MT}$, where $\delta_H = \gamma_G
\operatorname{tr}(\Sigma)+\sigma^2$. Using the above statistical characterization of $\mathbf{G}$ and the approximation of $\mathbf{K}_j$, we obtain $\|\boldsymbol{\mu}_{jk}\|_{\mathbf{K}_k^{-1}}^2\approx \|\boldsymbol{\mu}_{jk}\|_
{\mathbf{K}_k^{-1}
\mathbf{K}_j
\mathbf{K}_k^{-1}}^2\approx 
\left\|
\mu_{jk}
\right\|^2/\beta$, where $\beta=\delta_H/(MT\gamma_G)$. Adopting the above approximations in the MAP error bound in \eqref{errorTermNew}, we obtain the following expression, which does not require instantaneous channel or precoder information and is used as the feature-learning objective
\textcolor{black}{\begin{align}
\bar{P}_e
=
\sum_{j=1}^C p_j\sum_{k\neq j}  Q\left(
\frac{
\displaystyle
\left\|
\mu_{jk}
\right\|^2
+
\beta\log
\left(
{p_j}/{p_k}
\right)
}{
\displaystyle
\sqrt{
2\beta
\left\|
\mu_{jk}
\right\|^2
}
}
\right).
\label{eq_channel_independent_Q}
\end{align}
}
To improve robustness to different wireless operating conditions, $\beta$ is randomly selected from a predefined range for each mini-batch. This enables the feature extractor to learn representations that are robust to variations in the effective channel, precoding, and noise conditions rather than being tailored to a specific system configuration.

 The proposed objective has a computational complexity of $\mathcal{O}(DC^2)$, compared with $\mathcal{O}(BD^2+CD^3)$ for the MCR$^2$-based objective in~\cite{Cai2024MCR2}. Thus, the proposed objective has a lower asymptotic complexity, particularly for high-dimensional feature representations or large mini-batch sizes.

\subsection{Precoder Design}
\label{Section_prcoder}
To design the precoder matrix $\mathbf{V}$, we minimize \textcolor{black}{a tractable approximation of the MAP error bound} in \eqref{errorTermNew}. Specifically, we approximate the received-signal covariance matrices as diagonal and class-independent~\cite{ZhangFiniteFluid,Abadi2019Diffusion,Zhang2025Ahmadi,Ahmadi2024TWCIntegrated,Ahmadi2025Feedback}, i.e.,
$\mathbf{K}_i\approx\mathrm{diag}\{g_1(\mathbf{V}),g_2(\mathbf{V}),\ldots,g_{MT}(\mathbf{V})\}$, where $g_{l}(\mathbf{V})=\tilde{\mathbf{h}}_l\mathbf{V}\Sigma\mathbf{V}^{H}\tilde{\mathbf{h}}_l^{H}
+\sigma^2$, $\Sigma=\sum_{i=1}^Cp_i\Sigma_i$, and $\tilde{\mathbf{h}}_l$ denotes the $l$th row of $\tilde{\mathbf{H}}$. In addition, to obtain a tractable optimization problem, we approximate the Gaussian $Q$-function in \eqref{eq_pairwiseExpanded} using the exponential approximation
$Q(x)\approx\exp(-\tau x^2)$, where $\tau>0$ is a tuning parameter controlling the approximation accuracy. Accordingly, the precoder is obtained by solving
\begin{align}
  \min_{\mathbf{V}} \quad
  &\sum_{j=1}^C
    \sum_{k\neq j}p_j
    \mathcal{L}_{jk}(\mathbf{V}),\\\nonumber
  \text{s.t.}\quad
  &\mathbf{V}=\mathrm{blkdiag}(\mathbf{V}_1,\ldots,\mathbf{V}_K),\\\nonumber
  &\textcolor{black}{\mathbb{E}\{\|\mathbf{V}\mathbf{x}\|^2\}\leq PT},
\end{align}
with
\begin{align}
\mathcal{L}_{jk}(\mathbf V)
=
\exp\left(
-\tau
\frac{
\left(
S_{jk}(\mathbf V)
+
\log\left(\frac{p_j}{p_k}\right)
\right)^2
}{
2S_{jk}(\mathbf V)
}
\right),
\label{Cost_approximate}
\end{align}
where  $S_{jk}(\mathbf V)
=
\sum_{m=1}^{MT}
{
|\tilde{\mathbf h}_m\mathbf V\mu_{jk}|^2
}/{
g_m(\mathbf V)
}$, and the block-diagonal structure and dimensions of $\mathbf V$ and $\mathbf V_k$, $k=1,\ldots,K$, are given in Section~\ref{Secion_SysModel}.

To optimize the precoder, we employ an iterative covariance-update and projected gradient descent procedure. At the $t$th iteration, $g_m(\mathbf V)$ is first evaluated using the current precoder $\mathbf V^{(t)}$ and then held fixed during the gradient update. After the gradient step, the updated matrix is projected onto the set of block-diagonal matrices before being normalized to satisfy the transmit power constraint. Specifically, the update is given by    \begin{subequations}
    \begin{align}
\mathbf{V}_o^{(t+1)}&= 
\mathcal{B}\left(
\mathbf{V}^{(t)}\textcolor{black}{
-\eta}
\sum_{j=1}^{C}
\sum_{k\neq j}
p_j
\mathcal{L}_{jk}'
\left(\mathbf{V}^{(t)}\right)
\right),\\
\mathbf{V}^{(t+1)}
&=
\alpha_{\mathcal{P}}\mathbf{V}_o^{(t+1)},
\end{align}
\label{updateEquatoin}
\end{subequations}
where $\mathcal{B}(\cdot)$ enforces the block-diagonal structure of $\mathbf{V}$ by setting its off-diagonal blocks to zero, $\eta$ is the step size, $\alpha_{\mathcal P}$ is a normalization factor that rescales the updated precoder to satisfy the transmit power constraint, and $
\mathcal L_{jk}'
(\mathbf V)$ is the gradient of $
\mathcal L_{jk}
(\mathbf V)$ with respect to $\mathbf{V}$ which is 
\textcolor{black}{\begin{align}
\mathcal{L}_{jk}'(\mathbf V)
&=
-\frac{\tau}{2}
\mathcal{L}_{jk}(\mathbf V)
\left(
1-
\frac{
\log^2\left(p_j/p_k\right)
}{
S_{jk}^2(\mathbf V)
}
\right)
S_{jk}'(\mathbf V),
\label{eq_gradient_pairwise}
\end{align}
}
with 
\begin{align}
S_{jk}'(\mathbf V)
&=
2
\sum_{m=1}^{MT}
\frac{
\tilde{\mathbf h}_m^H
\tilde{\mathbf h}_m
\mathbf V
\mu_{jk}\mu_{jk}^H
}{
g_m(\mathbf V)
}.
\label{eq_S_gradient}
\end{align}
\textcolor{black}{Using the class-independent covariance matrix $\Sigma$, the average transmit power in \eqref{eq_powerConstraint} can be expressed as $
\mathbb{E}\{\|\mathbf V\mathbf x\|^2\}
=
\mathrm{tr}(\mathbf V\Sigma\mathbf V^H)
+\sum_{j=1}^{C}p_j
\|\mathbf V\mu_j\|^2$, which leads to
\begin{align}
\alpha_{\mathcal P}
=
\sqrt{
\frac{PT}
{
\mathrm{tr}\left(\mathbf{V}_o^{(t+1)}\Sigma{\mathbf{V}_o^{(t+1)}}^H\right)
+\sum_{j=1}^C p_j
\|\mathbf{V}_o^{(t+1)}\mu_j\|^2
}
}.
\label{alpha_p}
\end{align} 
}
As shown in Algorithm~\ref{alg_precoder}, the precoder matrix is updated using \eqref{updateEquatoin} over $I_0$ iterations, and the final updated matrix $\textcolor{black}{\mathbf{V}^{(I_0)}}$ is used as the precoding matrix for transmission.

The proposed precoder optimization has a per-iteration computational complexity
upper bounded in order by
$\mathcal{O}\!\left(TKX^2C^2+TK^2X^3\right)$, where $X=\max(M,N_k,D_k)$. In contrast, the MCR$^2$-based precoder in~\cite{Cai2024MCR2} has a per-iteration computational complexity of $\mathcal{O}\!\left(
D^3+C M^3 T^3+I_1 K T^3 N_k^3 D_k^3
\right)$, where $I_1$ denotes the number of iterations required for the bisection-based updates.

\begin{algorithm}
\caption{Precoder Optimization}
\label{alg_precoder}
\textcolor{black}{
{\small
\textbf{Input:} Channel matrix $\tilde{\mathbf H}$, class statistics $\{\mu_j,\Sigma_j\}_{j=1}^{C}$, initialized precoder $\mathbf V^{(0)}$, step size $\eta$, transmit power constraint $PT$.\\
\textbf{Output:} Optimized precoder $\mathbf V^{(I_0)}$.\\
\For{$t=0,1,\ldots,I_0-1$}
{
Evaluate $\mathcal{L}_{jk}(\mathbf V^{(t)})$ \& $\mathcal{L}_{jk}'(\mathbf V^{(t)})$ from \eqref{Cost_approximate} and \eqref{eq_gradient_pairwise}.\\
Update the precoder using \eqref{updateEquatoin} with $\alpha_{\mathcal P}$ defined in \eqref{alpha_p}.
}
}
}
\end{algorithm}

\begin{figure}[t!]
		\centering
		\includegraphics[width=.855\linewidth, trim=110 245 112
		252, clip]{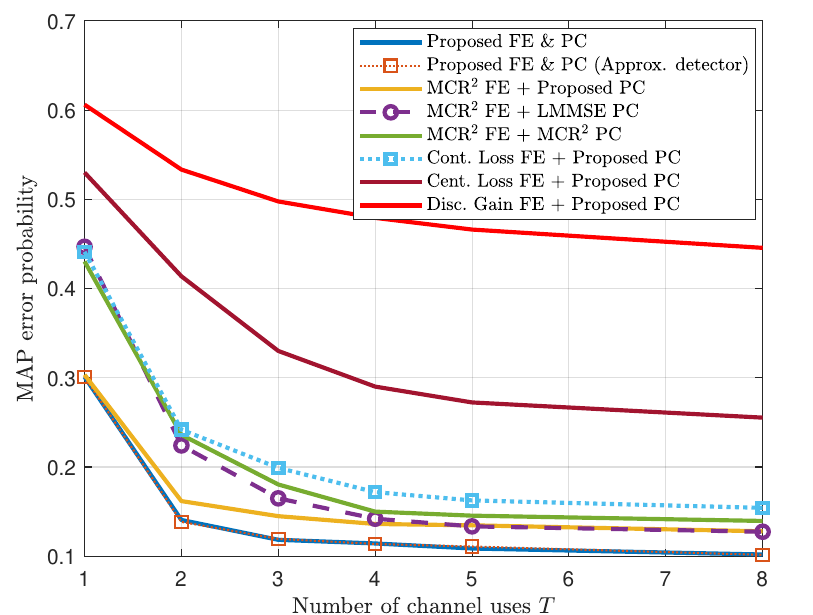}
		\caption{{\small MAP error probability versus the number of channel uses $T$.}}
		\label{fig:MAP_error_T}
	\end{figure}
    \begin{table}[t]
\centering
\caption{Average execution time (ms) of the considered precoders for different numbers of channel uses $T$.}
\label{tab:precoder_runtime}
\begin{tabular}{c|ccc}
\hline
$T$ & Proposed PC & LMMSE PC & MCR$^2$ PC \\
\hline
1 & 3.96 & 5.55 & 12.54 \\
2 & 4.04 & 5.94 & 20.82 \\
3 & 5.58 & 7.02 & 28.31 \\
4 & 6.18 & 7.31 & 41.98 \\
5 & 8.8 & 12.58 & 102.19 \\
8 & 17.06 & 20.61 & 153.21 \\
\hline
\end{tabular}
\end{table}

\begin{figure}[t!]
		\centering
		\includegraphics[width=.855\linewidth, trim=110 240 112
		252, clip]{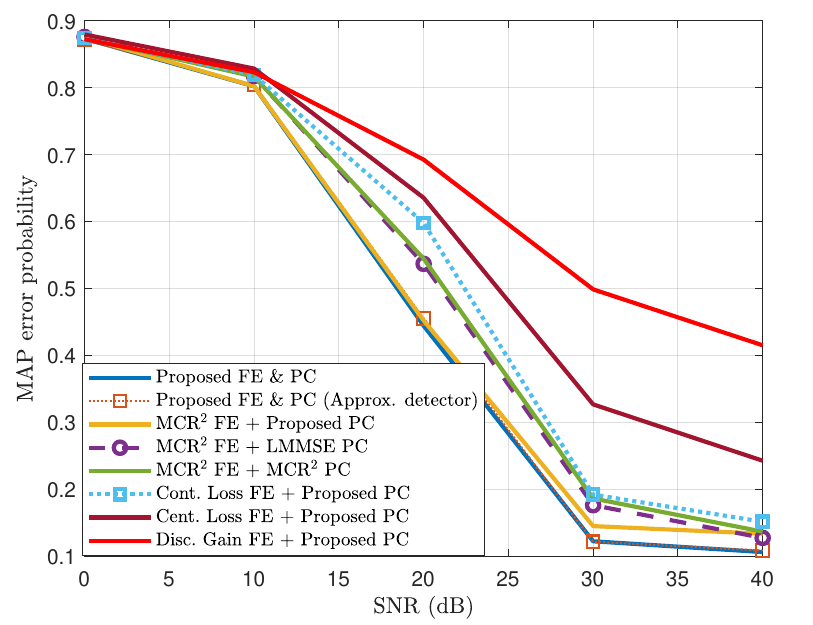}
\caption{{\small MAP error probability versus SNR.}}
\label{fig:MAP_error_SNR}
	\end{figure}
    

\section{Numerical Results}
\label{Sec_numResults}
In this section, we evaluate the performance of the proposed task-oriented communication framework. The wireless channel is modeled as a spatially correlated Rayleigh fading channel, where $\rho\in[0,1)$ denotes the spatial correlation parameter, with $\rho=0$ corresponding to independent Rayleigh fading. Unless stated otherwise, the following settings are used throughout the simulations. We set $\rho=0$, $P=1$ W ($30$ dBm), $\sigma^2=1$, $N_k=2$, $M=4$, $T=3$, $K=2$, and $D_k=4$, with uniform class priors. We consider the ModelNet10 dataset with $K$ views per sample, where each view is assigned to a dedicated worker. Each worker processes its assigned view using a ResNet-18 backbone initialized with ImageNet-pretrained weights, followed by a two-layer fully connected projection head that maps the 512-dimensional representation to a $2D_k$-dimensional latent representation. For feature extractor training, we use a batch size of $B=64$, a learning rate of $10^{-4}$, and 20 training epochs, with $\beta$ uniformly selected from $[3\times10^{-4},90]$ at each mini-batch. For the precoder optimization in Section~\ref{Section_prcoder}, we set $\eta=10$, $\tau=0.7$, and $I_0=10$. The feature extractor is implemented in Python 3.11.2 using PyTorch 2.1.2 with CUDA 12.0 and trained on an NVIDIA RTX A5000 GPU, whereas the precoder optimization is implemented in MATLAB R2024b and executed on an AMD Ryzen 7 PRO 7840U CPU with 30.67 GB of RAM.

\begin{figure}
		\centering
		\includegraphics[width=.854\linewidth, trim=110 240 112
		252, clip]{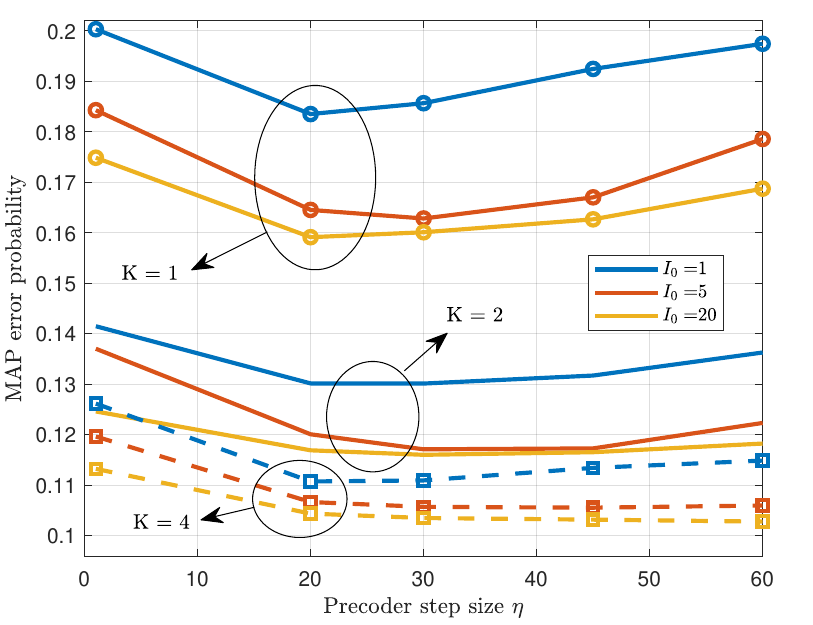}
\caption{{\small MAP error probability versus the precoder step size $\eta$ for the proposed FE and PC with different numbers of precoder iterations $I_0$ and workers $K$.}}
\label{fig:MAP_error_I0}
	\end{figure}
    
In the simulations, we consider different feature extractors (FEs) and precoders (PCs). The proposed FE and proposed PC denote the feature extractor and precoder developed in Sections~\ref{SectionNN} and~\ref{Section_prcoder}, respectively. The MCR$^2$ FE and MCR$^2$ PC denote the corresponding methods proposed in~\cite{Cai2024MCR2}, while the LMMSE PC denotes the precoder proposed in~\cite{LMMSE}, with the modifications described in Appendix~A of~\cite{Cai2024MCR2}. We also consider center loss (Cent.~Loss), contrastive loss (Cont.~Loss), and discriminant gain (Disc.~Gain) as alternative feature-learning objectives, where the latter maximizes
$\sum_{j=1}^{C}\sum_{k\neq j}p_j\mu_{jk}^{H}\Sigma^{-1}\mu_{jk}$. \textcolor{black}{Unless otherwise stated, the MAP detector in \eqref{MAP_general_form} is employed at the receiver for classification. To investigate the impact of the approximations made in Section~\ref{Section_prcoder}, we also consider an Approx. detector, which uses the same MAP detector in \eqref{MAP_general_form} but replaces the exact received covariance matrices with their diagonal and class-independent approximations.}

In Fig.~\ref{fig:MAP_error_T} and Fig.~\ref{fig:MAP_error_SNR}, we plot the MAP error probability as a function of $T$ and SNR ($\mathrm{SNR}=P/\sigma^2$), respectively, for different feature extractors and precoders. As shown in these figures, the proposed FE and PC combination consistently achieves the lowest MAP error probability across all considered values of $T$ and SNR. The experiments in Fig.~\ref{fig:MAP_error_T} are conducted with $\rho=0$, whereas Fig.~\ref{fig:MAP_error_SNR} considers a spatially correlated Rayleigh fading channel with $\rho=0.5$. As expected, increasing either $T$ or SNR improves the classification performance of all considered schemes. With the MCR$^2$ FE fixed, the comparison reflects the performance of the different precoders only, showing that the proposed precoder achieves lower MAP error probability than both the LMMSE and MCR$^2$ precoders, while the LMMSE and MCR$^2$ precoders exhibit similar performance, with a slight advantage for the LMMSE precoder. With the proposed PC fixed, the comparison reflects the performance of the different feature extractors only, showing that the proposed FE achieves lower MAP error probability than the MCR$^2$, Cont.~Loss, Cent.~Loss, and Disc.~Gain feature extractors, demonstrating the effectiveness of the proposed feature-learning objective. \textcolor{black}{Furthermore, for the proposed FE and PC combination, we observe that the approximated and exact MAP detectors achieve similar performance in both figures, indicating that the approximations introduced in Section~\ref{Section_prcoder} provide a good approximation under the considered simulation conditions.} To further assess the computational cost, execution times for different $T$ are reported in Table~\ref{tab:precoder_runtime}. The proposed precoder is up to approximately $11.5\times$ faster than MCR$^2$ and has comparable or lower execution time than LMMSE.

\begin{figure}
		\centering
		\includegraphics[width=.854\linewidth, trim=110 240 112
		252, clip]{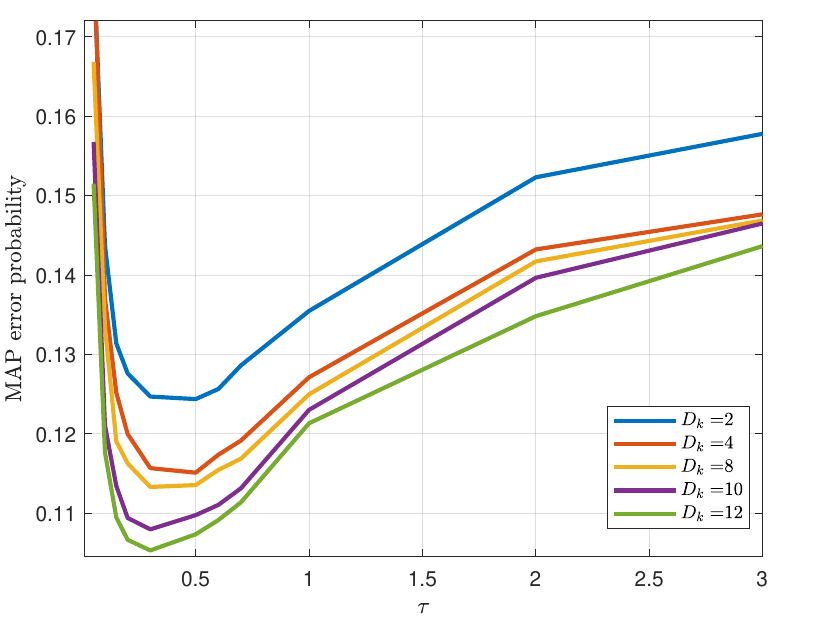}
\caption{{\small MAP error probability versus $\tau$, the parameter used to approximate the $Q$-function in Section~\ref{Section_prcoder}, for different feature lengths $D_k$ using the proposed FE and PC.}}
\label{fig:MAP_error_Dk}
	\end{figure}

In Fig.~\ref{fig:MAP_error_I0} and Fig.~\ref{fig:MAP_error_Dk}, we plot the MAP error probability as a function of the precoder step size $\eta$ and the tuning parameter $\tau$, respectively, for different values of $K$, $I_0$, and $D_k$. These figures show that choosing the step size $\eta$ or the design parameter $\tau$ either too large or too small can degrade the system performance, indicating the need for moderate values that provide a favorable performance. We also observe that increasing the number of workers $K$, feature length $D_k$, and number of precoder iterations $I_0$ generally improves the performance. However, these parameters also increase the computational complexity; therefore, practical system design requires a trade-off between classification performance and computational complexity.

    \balance
 \vspace{-3mm}

\section{Conclusion}
\label{SecConclusion}
In this paper, we proposed a MAP-driven task-oriented communication framework for distributed wireless classification. In particular, we derived feature-learning and precoding objectives using \textcolor{black}{tractable approximations of the MAP error probability.} Numerical results demonstrate improved classification accuracy and reduced computational complexity compared with existing approaches.

\end{document}